# Non-hysteretic branches inside the hysteresis loop in $VO_2$ films for focal plane array imaging bolometers

M. Gurvitch[a,b], S. Luryi[a,c], A. Polyakov[a], A. Shabalov[a]

[a]NY State Center for Advanced Sensor Technology (Sensor CAT),  [b]Dept. of Physics and Astronomy, [c]Dept. of Electrical and Computer Engineering
SUNY at Stony Brook, Stony Brook NY 11794

**Abstract**

In the resistive phase transition in $VO_2$, temperature excursions from points on the major hysteresis loop produce minor loops. We have found that for sufficiently small excursions these minor loops degenerate into single-valued, non-hysteretic branches (NHBs) linear in $\log(\rho)$ vs. T and having essentially the same or even higher temperature coefficient of resistance (TCR) as the semiconducting phase at room temperature. We explain this behavior and discuss the opportunities it presents for infrared imaging technology based on resistive microbolometers. It is possible to choose a NHB with $10^2$–$10^3$ times smaller resistivity than in a pure semiconducting phase, thus providing a microbolometer with low tunable resistivity and high TCR.

The prevailing implementation of the uncooled focal plane array (FPA) infrared imaging technology is based on resistive readout of individual $VO_x$[1-4] microbolometers. Initially, the pure-phase $VO_2$ was selected as the sensing material because of its above-room-temperature semiconductor-to-metal phase transition[5] in which the resistivity $\rho$ changes by a factor of ~$10^3$–$10^4$ in films, holding a promise of high bolometer responsivity. However, the hysteretic nature of the transition, the latent heat released/absorbed in the transition and the elevated temperatures of operation were considered undesirable,[2] so much so that initial attempts to use the phase transition were abandoned. A mixed vanadium oxide $VO_x$ with $x \approx 2$ was nevertheless retained as a semiconductor sensor material because it was found to posses at 25 C an attractive combination of reasonably high TCR = $(1/\rho)d\rho/dT$ and low $R_\square = \rho/d$, as well as moderately low 1/f noise[1-4]. The $VO_x$ films, typically 50 nm thick, provide room temperature TCR $\approx -2\%$ and $R_\square \sim 10 - 200$ k$\Omega$.[1-4] Practical FPA devices, however, utilize mainly $R_\square \sim 10 - 20$ k$\Omega$[3], as higher $R_\square$ causes problems in matching to the readout amplifiers, in sensor heating by the readout current, and in Johnson's noise, which is the prevailing noise source even at $R_\square = 20$ k$\Omega$[3]. The high resistance of pure-phase $VO_2$ films is actually the main reason why they are not used in FPA technology, not even at 25 C where the resistivity $\rho$ of pure phase ranges from 0.1 $\Omega$m to 1.0 $\Omega$m.[6] These values of $\rho$ correspond to $R_\square = 2 - 20$ M$\Omega$ in a 50 nm film, which is $10^2 - 10^3$ times (!) higher than required.

If it were not for this large resistance mismatch, pure phase $VO_2$ would be preferred over $VO_x$ for the following reasons:

1. Bolometer responsivity and signal-to-noise ratio are proportional to TCR,[7] so higher TCR is essential to bolometer performance. In $VO_2$, TCR is from $-3\%$ to $-5\%$[6] which is significantly higher than $-2\%$ in $VO_x$.

2. A well-defined single phase sensor material allows for an easier process control compared to a need to reproduce and make uniform layers of a mixed, ill-defined, ill-behaved $VO_x$.

3. A pure phase sensor material with fewer defects should have a lower 1/f noise.

We list these attractive features here because we have discovered a rather remarkable new phenomenon that offers the possibility of preserving the high TCR while dramatically, by orders

4of magnitude, lowering the $R_\square$ in pure phase $VO_2$. Moreover, our explanation of the new phenomenon indicates that its use circumvents the other difficulties associated with the phase transition (the hysteretic behavior, the latent heat and the excess noise in the two-phase region).

Our polycrystalline $VO_2$ films were deposited by POP[6] at 400 C and by PLD[8] at 600 - 650 C on oxidized silicon and sapphire substrates. The resistance ratio $\rho_s(25\ C)/\rho_m(90\ C)$ was $2\times10^3$ in the best POP films, while in the PLD films it reached $5\times10^3$ on $Si/SiO_2$ and $4\times10^4$ on sapphire. The four-probe $\rho(T)$ measurements were performed with 0.4–0.8 µA measuring current, in a thermostat with temperature sweep rate of 0.5 C/min. The data were collected with the use of an automated measuring system.

We studied minor nested loops "attached" to the major hysteresis loop in the resistivity of $VO_2$ films. These minor loops can be produced starting from any *attachment* temperature $T_0$ on the major loop by making a *backward round-trip excursion* from that temperature. For $T_0$ on the heating branch (HB) the roundtrip excursion denotes a $T_0 \to T_0–\Delta T \to T_0$ process, i.e. cooling down from $T_0$ to $T_0–\Delta T$ (we shall refer to $\Delta T > 0$ as the *excursion length*) and then warming up by $\Delta T$ back to $T_0$. On the cooling branch (CB) the backward direction is that of warming up, and so the roundtrip excursion is $T_0 \to T_0+\Delta T \to T_0$. Excursions in the opposite (forward) direction do not produce a minor loop.

Figure 1 shows some of the minor loops traced in backward round-trip excursions with $\Delta T$ ranging from ∼ 2 C to ∼ 25 C. Big and small excursions are shown in Fig. 1a. Studying minor loops with progressively smaller excursions $\Delta T$, we discovered a threshold $\Delta T^* \sim 4$ C below which there is a qualitative change in behavior. We have found that for $\Delta T < \Delta T^*$ the minor loop flattens out and becomes a single-valued branch linear in $\log(\rho)$ vs. T. This is what we call a non-hysteretic branch (NHB). A NHB can be initiated from any attachment point on the major loop, either on the HB or CB. The minor loop data are removed in Fig. 1b, which shows the major loop with NHBs only. We traced these NHBs back and forth within $\Delta T$, plotting them on an expanded scale, and confirmed that they were both linear in $\log(\rho)$ vs. T. to the precision of our measurements and entirely non-hysteretic (single valued).

An eye examination of Fig. 1b suggests that, except for the region close to the high-temperature merging point of the major loop $T_M \sim 90$ C, all NHBs exhibit similar slopes on the $\log(\rho)$ vs. T plot, or similar TCR = 2.3 d[$\log(\rho)$]/dT, and that these slopes are also similar to the slope in the pure semiconductor (S) phase below the phase transition. Closer examination reveals that the NHB slopes are actually higher than the S-phase slope. Exact fitting shows that the S-phase slope near 25±5 C corresponds to TCR = −(2.5 ± 0.25)% , whereas for various NHBs the TCR ranges from −2.9% to −4.0%. This excludes a region close to $T_M \approx 90C$, where TCR = −(1.2± 0.1) %.

Although only PLD results are presented in Fig. 1, the POP films show similar behavior.

So long as we do not exceed the critical excursion length, the TCR on each NHB remains stable and reproducible after multiple cyclings. This includes repeated excursions back and forth about a mid-point of a given NHB. We can also go out on a global trip over the entire major loop, come back to the same attachment point $T_0$ and then effect a small backward excursion. This results in the same TCR which is therefore characteristic of a given NHB ($T_0$).

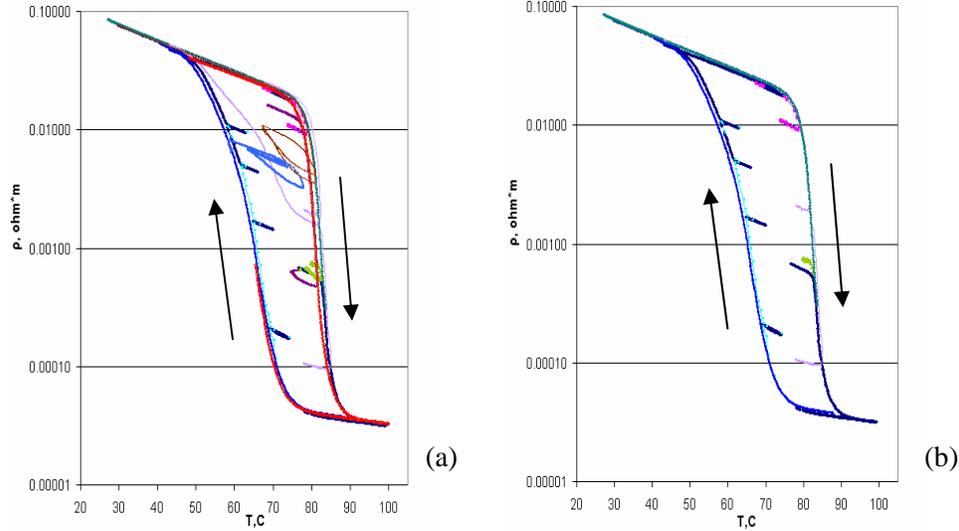

**Fig.1 (a)** Minor loops of various lengths inside the major hysteresis loop in a 95 nm VO$_2$ film deposited by PLD on SiO$_2$/Si; the resistance ratio $\rho_s(25\,C)/\rho_m(90\,C) = 2,300$. The arrows indicate HB and CB. Small loops become single valued, non-hysteretic branches (NHB) while essentially maintaining the TCR of the semiconductor phase. **(b)** The same sample with minor loop data removed, showing only the NHBs.

Let us now give a qualitative explanation to the observed NHB phenomenon. We wish to understand both the non-hysteretic behavior and the fact that all NHBs have similar TCR that is essentially the S-phase TCR but slightly higher.

The hysteretic region in VO$_2$ is a mixed state consisting of both the semiconductor (S) and the metallic (M) phase regions. Each such region located around a point $(x,y)$ transitions into the other phase at its own temperature $T_C(x,y)$ with an intrinsic hysteresis characterized by the coercive temperature $T^*(x,y)$. In a macroscopic sample these parameters are continuously distributed. Ignoring for the sake of simplicity the variation in $T^*$, we shall assume that the film is characterized by a local $T_C(x,y)$. At a given temperature T inside the hysteretic loop, some parts of the film have $T_C(x,y) < T$ and some $T_C(x,y) > T$. In the first approximation, the boundary wall between the S and M phases is determined by the condition $T_C(x,y) = T$. In this approximation, the wall is highly irregular and its ruggedness corresponds to the scale at which one can define the local $T_C(x,y)$. On closer inspection, however, we need a refinement that takes into account the *boundary energy*, associated with the phase domain wall itself. The boundary energy is positive and to minimize its contribution to the free energy the domain walls are relatively smooth.

For concreteness, let us consider the heating branch. As the temperature rises, the area of the M phase increases. Let us focus on two metallic lakes that are about to merge. Since the boundary is smooth, at some temperature the distance between the lakes becomes smaller than the radius of curvature of either lake at the point they will eventually touch. In this situation simple geometric considerations show that at some $T = T_{cr}$ the following two configurations will have equal energies: one comprising two disconnected M phase lakes that are near touching, but not quite, and the other with a finite link formed between the two lakes. Both configurations are characterized by equal boundary lengths and therefore have equal free energy. In the thermodynamic sense one could call the $T_{cr}$ the critical temperature for the link formation, if we could wait long enough. The actual transition forming a local link, however, does not occur at that temperature because of an immense *kinetic* barrier between these two macroscopically different configurations. The transition occurs at a higher $T_0 = T + \Delta T$ when it is actually forced, i.e. when the two phases touch at a point. We associate the steep slopes of the major loop with the quasi-continuous formation of such links, i.e. with local topological changes. On the HB the steep slope is associated with the merger of metallic lakes, on the CB it is the disconnection of metallic links which in 2D is the same thing as the linkage of semiconductor regions. The above-mentioned point $T_M$ is associated with the establishment of an infinite metallic cluster (percolation transition).

Consider now a small excursion backwards from $T_0$. As the temperature decreases, the last formed M-link does not disappear immediately for the same kinetic reason. One has two S regions that need to touch in order to wipe out the M-link. It takes a backward excursion of amplitude $\Delta T$ to establish an S-link and thus disconnect the last M-link. So long as we are within $\Delta T$, i.e. stay on the same NHB, the area of S and M domains changes continuously, but the topology is stable and no new links are formed. Within the range of that stable "frozen" topology, the resistivity of NHB is predominantly controlled by the percolating semiconductor phase. The slope (TCR) on the NHB may be slightly higher than that of the semiconductor phase itself because it includes not only the temperature variation of semiconductor resistivity but also the smooth change of geometry.

Similar consideration can be applied to the cooling branch of the major loop.

As long as the S-phase forms a global cluster (and therefore the M-phase is disconnected), the S-phase TCR will be observed in the NHBs. Once the M-phase percolates, it shorts out the S-phase resistivity. This is what is observed near $T_M \approx 90$ C (Fig. 1). There is a corresponding point near the upturn of the cooling curve, at about 75 C in Fig. 1, where S phase again forms the infinite semiconductor cluster. In a 2D film these points would correspond to 1:1 ratio of phase areas; in a real film of finite thickness both phases can percolate in some range of temperatures.

The discovered NHB phenomenon can be employed in FPA applications. One can now make use of a $VO_2$ film to fabricate the pixellated bolometric sensor array. The deposition process for $VO_2$ is compatible with the normal bolometer fabrication process: using POP we deposited good $VO_2$ at below 400 C.[6] The sensor array should be set to operate within a NHB attached either to the HB (e.g. at 85 C in Fig. 1) or to the CB (e.g. at 70 C in Fig. 1). The operating temperature $T_{OP}$ (i.e. the temperature at which the sensor array is stabilized awaiting the projected IR signal) should be removed from either end of an NHB, for example, it can be chosen at the mid-point. In this case, the available range is between $T_{OP} - \Delta T^*/2$ and $T_{OP} + \Delta T^*/2$. The narrow operation range of a few degrees should not present a problem in IR visualization, where much smaller temperature changes are typically detected. Positioning an array at the operating temperature will require:

- On a HB, warming up to $T_0$ and cooling down to $T_{OP} = T_0 - \Delta T^*/2$.

- On a CB, warming up to above $T_M$, cooling down to $T_0$, and warming up to $T_{OP} = T_0 + \Delta T^*/2$.

The sensing layer will have high TCR and at the same time the chosen (i.e. tunable) resistivity 2−3 orders of magnitude lower than the room temperature (semiconductor) value. This will lower the Johnson noise voltage by a factor of ~10−30. The tunable $R_\square$ can be used to match the bolometer resistance to the input impedance of the front-end amplifier. We further expect that, because of the frozen topology within a NHB, there will be no 1/f noise contribution associated with processes of random link formation. Since all processes on a given NHB are of continuous nature, the release and absorption of the latent heat of the phase transition should also be insignificant in small temperature excursions within a NHB.

In conclusion, we have discovered a new phenomenon associated with the semiconductor-metal transition in VO$_2$ films, characterized by the appearance of stable and reproducible non-hysteretic branches off the main transition loop. We have qualitatively explained the effect by an argument involving suppression of the local topological changes by the boundary energy. The effect can be useful for focal plane array infrared imaging systems. It enables a bolometric element that combines the high TCR with a low and tunable resistivity. The non-hysteretic branches are characterized by continuous variations of local phase areas and therefore their use will circumvent the unwelcome effects usually associated with phase-transition based devices.

**Acknowledgements:** We thank Dr. David Westerfeld for developing an automated resistivity measurement system and Dr. Arsen Subashiev for useful theoretical discussions. This work was partially supported by the New York State Foundation for Science, Technology and Innovation (NYSTAR) through the facilities of the Center for Advanced Sensor Technology at Stony Brook. We thank Dr. Peter Shkolnikov for his continuing interest and support of this work.